\documentclass[journal=jacsat,manuscript=article]{achemso}
\usepackage[version=3]{mhchem} % Formula subscripts using \ce{}

\usepackage{parskip}
\usepackage{lineno}
\usepackage{amsthm}

\theoremstyle{plain}

\theoremstyle{definition}

\theoremstyle{remark}

\usepackage{hyperref}
\usepackage{comment}

\title{Correlative Ultrafast Imaging of a Propagating Photo-Driven Phase Transition Using 4D STEM}

\author{
    Arthur Niedermayr
}
\affiliation{Department of Materials and Nano Physics, School of Engineering Sciences, KTH Royal Institute of Technology, Stockholm SE-100 44, Sweden.}
\email{arthurn@kth.se}

\author{
    Jianyu Wu
}
\affiliation{Department of Materials and Nano Physics, School of Engineering Sciences, KTH Royal Institute of Technology, Stockholm SE-100 44, Sweden.}
\author{
    Bertina Fisher
}
\affiliation{Department of Physics, Technion--Israel Institute of Technology, Haifa 32000, Israel.}
\author{
    Ido Kaminer
}
\affiliation{Department of Electrical and Computer Engineering, Technion--Israel Institute of Technology, Haifa 32000, Israel.}
\author{
    Jonas Weissenrieder
}
\affiliation{Department of Materials and Nano Physics, School of Engineering Sciences, KTH Royal Institute of Technology, Stockholm SE-100 44, Sweden.}
\email{jonas@kth.se}


\begin{document}

\begin{abstract}
Oxides exhibiting insulator-metal transitions are promising candidates for next-generation ultrafast electronic switching devices. However, critical gaps remain in understanding the onset of strain and its dynamics as these materials undergo structural transitions, particularly in nanostructured configurations. Here, we present ultrafast four-dimensional scanning transmission electron microscopy enabling virtual imaging and strain mapping at every point in space and time. Using this technique, we directly probe a laser-excited phase transition in the prototypical material vanadium dioxide (VO$_2$), recording its spatiotemporal propagation. This direct imaging capability reveals the dynamics of the structural phase transition and connects it to the resulting strain formation on picosecond timescales. This correlation reveals how atomic-scale symmetry breaking inherently generates lattice distortions, which then propagate to govern macroscopic property changes. Our findings provide new insights into the coupling between electronic, structural, and mechanical responses in correlated oxides under non-equilibrium conditions.

\begin{figure}[htbp]%
\centering
\includegraphics{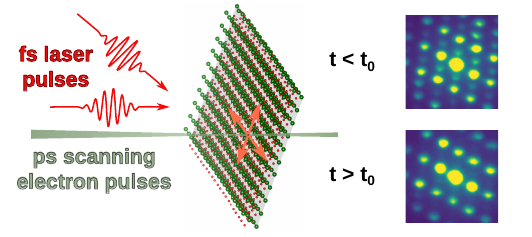}
\end{figure}

\end{abstract}

\maketitle
Photo-driven insulator-to-metal phase transitions in strongly correlated oxides hold significant potential for next-generation solid-state electronic devices \cite{mitrano_possible_2016, samizadeh_nikoo_electrical_2022,ohkoshi_synthesis_2010}, particularly for ultrafast switching applications. Exploiting the unique properties of these materials, such as their ability to rapidly transition between insulating and metallic states under light excitation, opens the possibility to design devices that operate on picosecond timescales. Their tunable optical and electrical properties make them highly attractive for applications in Mott field-effect transistors, memristive devices, thermal sensors, and chemical sensors \cite{yang_oxide_2011,gu_current-driven_2007, kim_temperature_2007, dicken_frequency_2009,driscoll_phase-transition_2009,ruzmetov_electrical_2009,strelcov_gas_2009,briggs_compact_2010,ruzmetov_three-terminal_2010,sood_universal_2021}. Additionally, the significant changes in reflectivity and conductivity associated with these phase transitions enable the development of ultrafast optical switches and adaptive electronic components. \\

Photo-excitation with short laser pulses, at high intensities, has in recent years become  a central method for studying the underlying behavior of strongly correlated oxides and of other quantum materials. Such short and intense pulse excitations inevitably leads to strain formation \cite{bertoni_elastically_2016}. Understanding the mechanism of strain formation and their influence on phase transitions is a long-standing challenge~\cite{park_measurement_2013, johnson_ultrafast_2022, ocallahan_inhomogeneity_2015}. Strain can locally modify the lattice structure, directly affecting the temperature, kinetics, and spatial inhomogeneity of the metal-insulator transition, making nanoscale strain mapping crucial to fully understand and control phase behavior. This is particularly important in thin films, where faster switching can be achieved compared to bulk counterparts, and where a larger fraction of the material interacts with the light due to reduced screening.\\

So far, most efforts to understand strain on ultrafast timescales have focused on the generation of acoustic waves, which lead to local sample bending and produce strong contrast. This mechanism has been extensively studied over the past few decades due to its relative ease of detection using bright- and dark-field imaging (or a combination thereof) with local diffractive probing~\cite{mckenna_spatiotemporal_2017,zhang_observation_2019,du_imaging_2020, zhang_imaging_2021,ji_influence_2023, wu_spatiotemporal_2024}. However, microscopic, local strain dynamics associated with structural phase transitions remain challenging to access and quantify. Despite ongoing efforts, open questions remain regarding the interplay between the structural phase transition and strain on nanometer length scales.

This work explores the mechanisms underlying structural phase transitions induced by pulsed optical excitation. The evolving phase transition generates local strain fields, which propagate through the material and dynamically reshape the energetic landscape, feeding back into the transition process. To address these limitations and deepen our understanding of the microscopic interplay between strain and phase transitions, we employ direct, quantitative imaging of the spatiotemporal evolution of both strain and structural phase separation. Using the seminal material vanadium dioxide (VO$_2$) as a model system, we demonstrate how this dual imaging approach reveals a correlation between structural distortions and phase transitions on ultrafast timescales.\\

It is well established that strain affects the transition temperature of VO$_2$ and can be harnessed to tune its electronic and structural properties \cite{park_measurement_2013,huber_ultrafast_2016}. Therefore, understanding how acoustic waves and other forms of strain influence performance is key for practical device applications. Notably, strain itself can act as a trigger for the phase transition \cite{donges_ultrafast_2016}. It thus remains a central challenge in the field to distinguish between optically and strain-induced mechanisms in the dynamics of VO$_2$. \\

To address this challenge, we track the intricate evolution of strain during optically induced phase transitions, offering new perspectives on the interplay between electronic excitation, structural changes, and strain at picosecond-nanometer resolution. A key innovation in our approach is the use of a spatially structured optical excitation, which not only induces electronic and structural phase transitions but also provides a means to probe them with high spatiotemporal resolution. The resulting patterned phase separation - characterized by periodic metallic and insulating domains - functions beyond the role of a conventional grating.  It enables two critical advancements: Firstly, it introduces optical variations on the scale of hundreds of nanometers, allowing access to dynamic processes at the combined picosecond-nanometer scale. Secondly, it imposes a well-defined excitation geometry, providing a reproducible framework to resolve spatial heterogeneity and isolate concurrent dynamic processes. This structured excitation opens up new pathways for precision studies of correlated materials and highlights potential applications in nanoscale optical devices, including energy-efficient transmission gratings for wavelengths below the material’s band gap.\\

Ultrafast strain dynamics are typically analyzed using ultrafast transmission electron microscopes \cite{zewail_4d_2006, kim_high-resolution_2024}, where bright- and dark-field imaging are the most commonly employed techniques \cite{mckenna_spatiotemporal_2017,zhang_observation_2019, zhang_imaging_2021}. However, these approaches are limited in that they primarily detect strain components aligned with the beam propagation direction due to the strong influence of bending and tilting on the local diffraction conditions. Furthermore, achieving quantitative, component-resolved strain mapping on ultrafast timescales remains an open challenge. In contrast, ultrafast four-dimensional scanning transmission electron microscopy (U-4D STEM) represents a powerful approach for quantitative strain determination. A few pioneering studies employed a variant of this technique: utilizing  convergent-beam electron diffraction (CBED)~\cite{ feist_nanoscale_2018, nakamura_visualizing_2022, shimojima_development_2023},  a technique that does not allow for simultaneous Bragg-resolved bright- or dark-field imaging. These studies have focused on acoustic strain wave analysis in well-established single-element materials, where strain gradients produce high-contrast signals \cite{feist_nanoscale_2018, nakamura_visualizing_2022, shimojima_development_2023}. CBED studies often rely on detailed calculations to interpret the CBED patterns for quantitative strain analysis \cite{rozeveld_determination_1993, ophus_four-dimensional_2019}. These calculations are necessary because CBED patterns are sensitive to the crystallographic orientation and state of strain in the sample, and the interpretation of these patterns requires complex modeling to extract accurate strain information. While CBED is an important technique for quantitative strain analysis, it is not straightforward to simultaneously track the evolving structural phase during ultrafast transitions.

While previous ultrafast diffraction studies employed CBED with comparatively large convergence angles, here we implement ultrafast 4D STEM using a nano-beam electron diffraction (NBED) configuration with a small convergence angle, enabling quantitative, component-resolved strain mapping on the nanometer scale. Although the distinction between CBED and NBED is gradual rather than sharp, to our knowledge the combination of ultrafast strain analysis and Bragg-resolved virtual imaging has never been demonstrated before. NBED also offers the additional advantage of simultaneous post-selection for various imaging modes, such as virtual bright-field, dark-field, or Z-contrast imaging, and the strain interpretation is more straightforward than in CBED. This simultaneous access to quantitative strain and Bragg-resolved imaging was not readily achievable in earlier ultrafast CBED studies.

\begin{figure}[htbp]%
\centering
\includegraphics[width=0.95\textwidth]{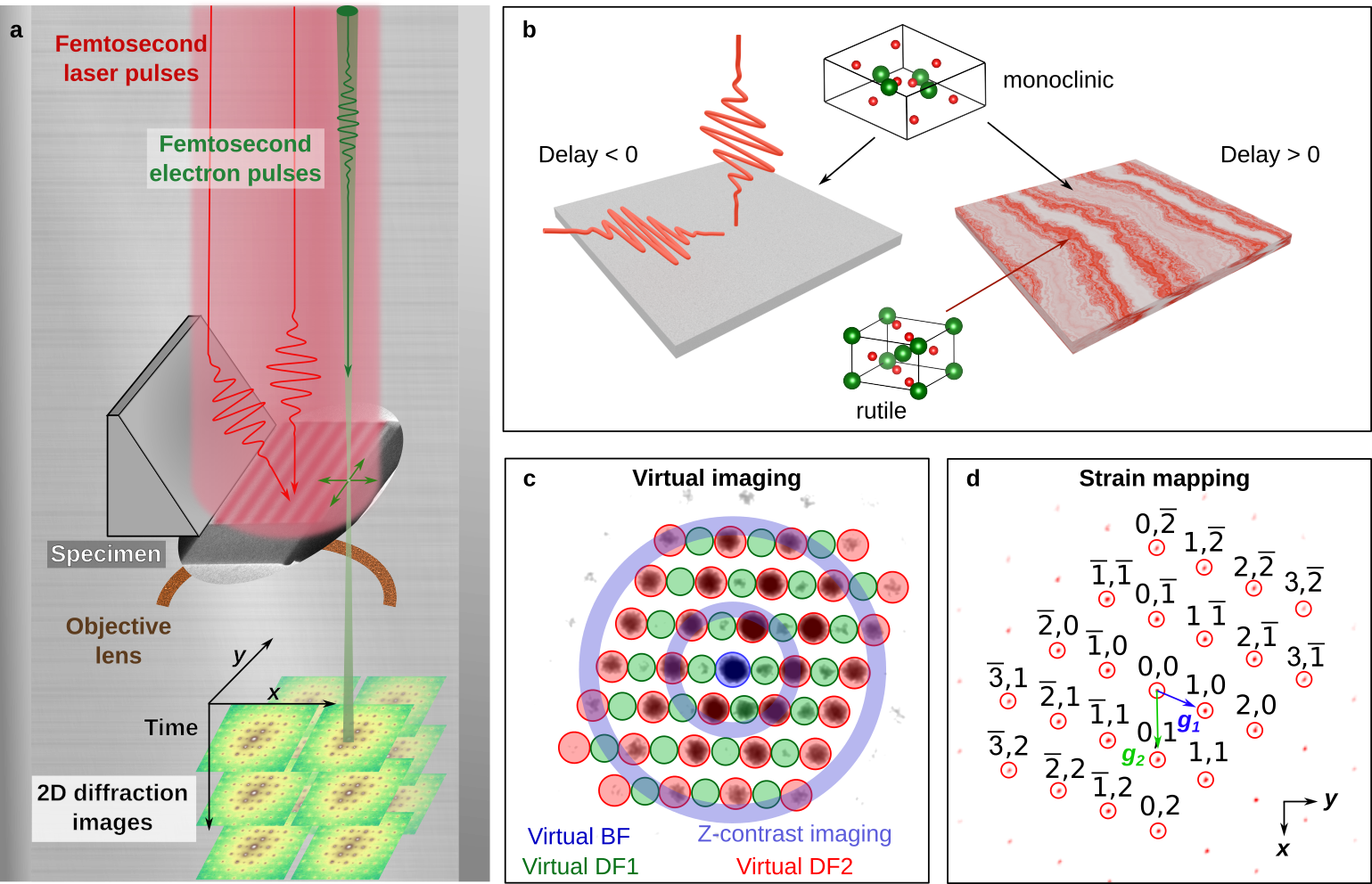}
\caption{\textbf{Ultrafast 4D STEM experiments. }(a) Schematic of the experimental setup: femtosecond laser pulses induce a patterned structural phase transition in a vanadium dioxide lamella, which is locally probed in real space using delayed femtosecond electron pulses. (b) A transient laser grating triggers a structural phase transition, resulting in different lattice structures in separate domains of the same lamella. (c) Arbitrary virtual masks enable the generation of ultrafast virtual images. (d) Ultrafast strain mapping captures the propagation dynamics of the structural phase transition by measuring the distance between multiple diffraction spots as a function of position on the specimen.} \label{fig1}
\end{figure}

Our experimental setup is shown in Fig. \ref{fig1}a. In a pump-probe configuration, a pulsed laser is reflected off a metallic mirror \cite{cao_femtosecond_2021} and overlaps with the direct beam on the specimen, generating a transient optical grating through self-interference. A delayed electron pulse is rastered across the specimen, generating a local diffraction pattern at each probe position. This approach enables investigation of the local structural response to a well-defined spatially structured excitation within an otherwise uniform specimen. The specimen’s response to a spatially patterned laser pulse is shown in Fig. \ref{fig1}b. Photon absorption in vanadium dioxide induces an electronic and structural phase transition, where illuminated regions of the sample transition from the monoclinic (M1) phase to the rutile (R) phase \cite{baum_4d_2007}. The combination of the structural phase transition, the heat gradient, and the resulting strain generates stress gradients and a strain mismatch. These effects induce a mechanical response in the thin, flexible lamella, resulting in its bending to accommodate the gradients. Our ultrafast 4D STEM approach enables virtual imaging through post-selection of masks in diffraction space (Fig. \ref{fig1}c) while also providing quantitative strain measurements through analysis of the diffraction patterns (Fig. \ref{fig1}d) \cite{savitzky_py4dstem_2021}. The determination and calibration of the temporal overlap between the optical pump and electron probe (time zero) are described in detail in the SI.
\\

\section*{Results}
\subsection*{Photo-driven ultrafast structural phase transition}
\begin{figure}[htbp]%
\centering
\includegraphics[width=0.95\textwidth]{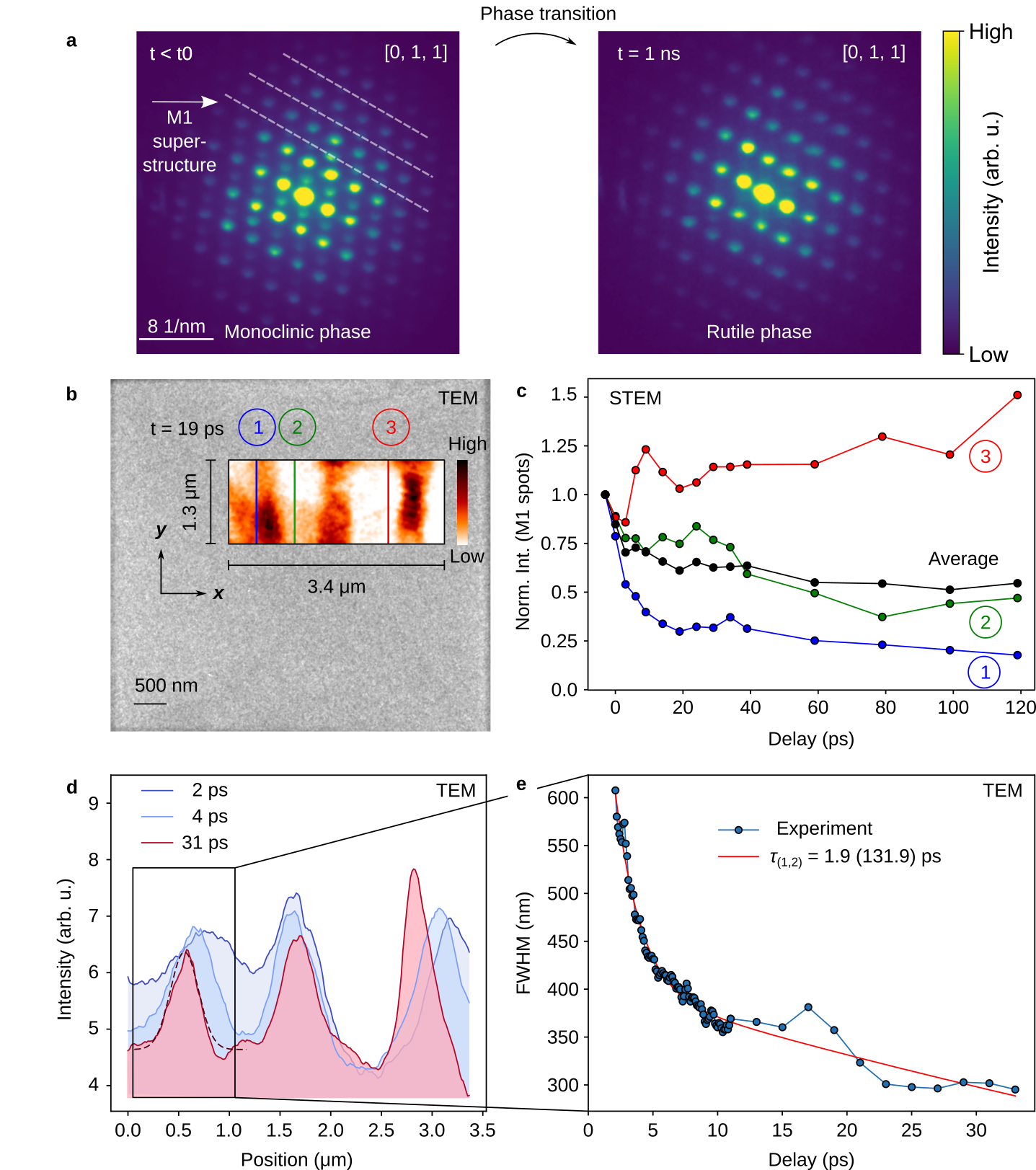}
\caption{\textbf{Transient grating excitation in vanadium dioxide on ultrafast timescales.} (a) Ultrafast electron diffraction patterns before and after the structural phase transition. The weak diffraction spots corresponding to the M1 superstructure, which appear at positions in between the main rutile reflections (effectively on every second line of the diffraction pattern), disappear upon completion of the transition to the rutile phase at 1~ns. (b) Ultrafast bright-field imaging following patterned light excitation. The background shows a static TEM image of the lamella. (c) Normalized intensity of the M1-exclusive spots, integrated over the line profiles of 4D diffraction scans performed in the highlighted window in (b). (d) Line profiles of the ultrafast bright-field images in (b) at different time-delays. (e) FWHM of Gaussian fits to the line profiles in (d) as a function of the pump-probe delay.}\label{fig2}
\end{figure}

The diffraction patterns corresponding to the monoclinic (M1) and rutile (R) phases are shown in Fig.~\ref{fig2}a. The M1 superstructure is characteristic of the monoclinic phase and is absent in the rutile phase, whereas the stronger diffraction spots are common to both phases. Bright-field imaging of the VO$_2$ lamella shows  the laser-induced grating (Fig. \ref{fig2}b), however fails to provide quantitative information about the specimen. To overcome this limitation, we perform additional 4D STEM on the indicated window in Fig. \ref{fig2}b and extract quantitative information (Fig.~\ref{fig2}c). This allows us to track the structural phase transition in space and time and quantify phase coexistence by diffraction analysis. In regions close to maximum interference of the optical grating, we observe a drop of the M1-phase–exclusive Bragg spots by 75\%, while in other regions the monoclinic phase appears slightly enhanced. The drop in intensity of the M1 spots reflects the local progression of the structural phase transition from the monoclinic to the rutile phase. Intensity variations exceeding the normalized value of 1 (red line in Fig.~\ref{fig2}c) arise from local sample bending and tilts, which modify the intersection of the reciprocal lattice with the Ewald sphere and thereby alter the diffraction conditions. Figures \ref{fig2}d,e show line profiles and corresponding analysis of the grating peak width as a function of the pump-probe delay. While bright-field contrast does not provide a quantitative measure of either strain or phase fraction, it serves as a sensitive real-space indicator of the overall photo-induced response. In particular, the evolution of the grating profile allows us to qualitatively follow the spatiotemporal redistribution of contrast on ultrafast timescales. From this analysis, we identify two characteristic timescales. The first time constant, approximately 2~ps, is consistent with the timescale of the photo-induced structural phase transition reported in previous ultrafast diffraction studies~\cite{baum_4d_2007}. The second, slower timescale of about 130~ps reflects the gradual build-up and real-space evolution of the grating contrast following excitation. The characteristic time constants are convoluted with the electron beam pulse duration, which is approximately 1.5 ps, as determined through photon-induced near-field electron microscopy. Importantly, the bright-field contrast does not directly track the evolution of the structural phase transition. The bright-field contrast reflects a convolution of multiple contributions, including diffuse scattering, transient lattice disorder, and local bending or tilts. These effects respond differently to optical excitation and are not phase-locked to the structural order parameter. Consequently, while bright-field imaging captures the overall spatial response to photo-excitation, the direct identification and tracking of the phase transition requires Bragg-resolved diffraction and quantitative strain analysis, as provided by ultrafast 4D STEM.

\subsection*{Ultrafast virtual imaging and contrast enhancement}
\begin{figure}[htbp]%
\centering
\includegraphics[width=0.9\textwidth]{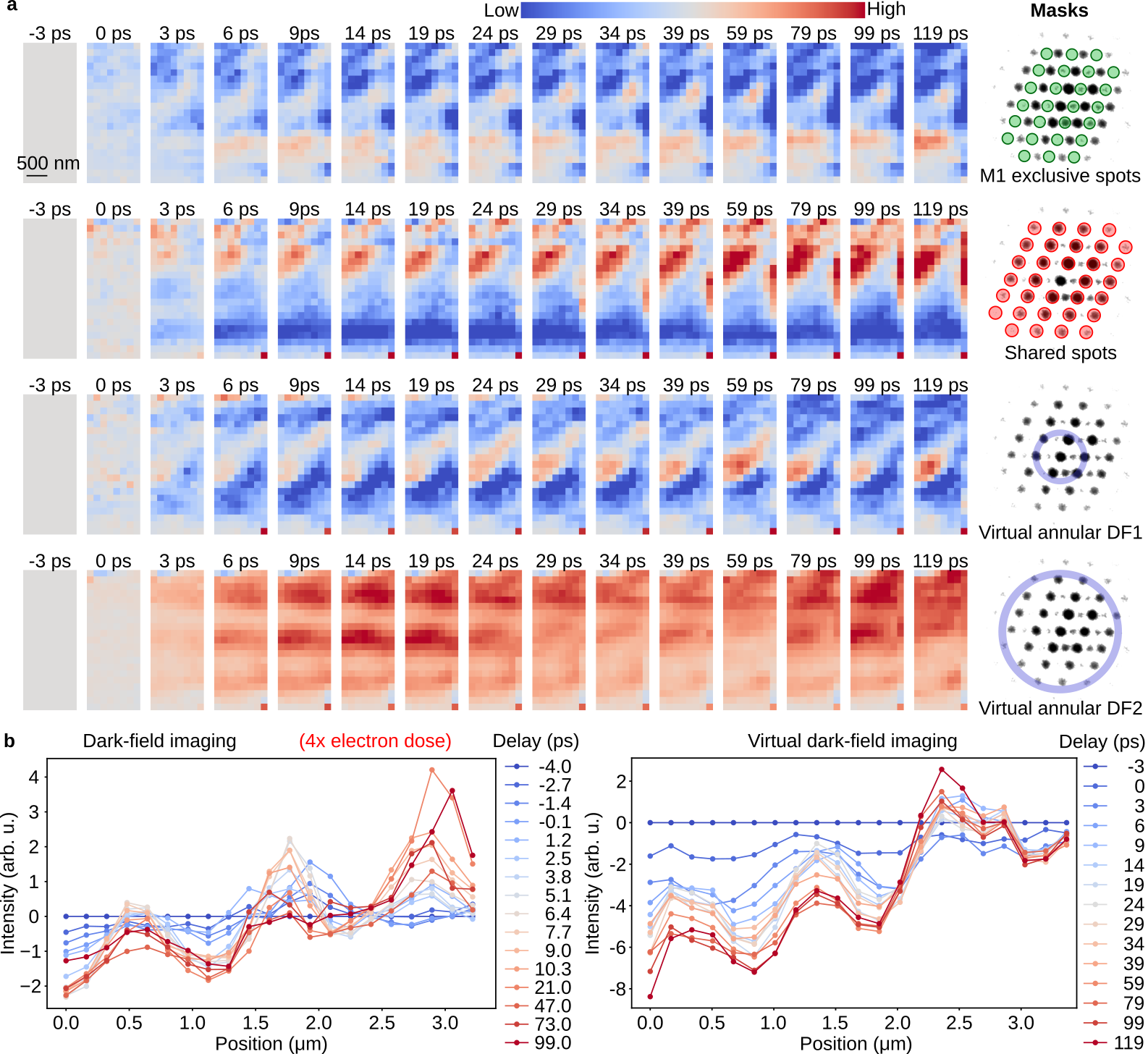}
\caption{\textbf{Dark-field imaging of a propagating photo-driven phase transition. }(a) Ultrafast diffraction-contrast imaging using various virtual masks (right panel) applied to the region shown in Fig. \ref{fig2}b. Different virtual masks highlight distinct specimen properties. For instance, specific masks can enhance the contrast of weak reflections or provide varying Z-contrast by selecting different scattering angles. (b) Comparison of line profiles from dark-field imaging using a single physical aperture (5 $\mu$m diameter) covering a single M1-exclusive spot versus virtual dark-field imaging incorporating all M1-exclusive spots, highlighting the enhanced contrast achieved in the latter.}\label{fig3}
\end{figure}

We employ the unique capabilities of ultrafast 4D STEM and apply several virtual dark-field masks which give us complementing virtual imaging capabilities.  The first panel of Fig. \ref{fig3}a illustrates the propagation of the structural phase transition by including the M1-exclusive Bragg spots through virtual imaging. As the material transitions from the M1 to the R phase, the intensity of these M1-exclusive spots is strongly suppressed (Fig. \ref{fig2}c) resulting in a corresponding decrease in the intensity of the virtual images at regions of high local fluence. Panel 2,3,4 show complementary information, by unmasking the shared spots of the monoclinic and rutile phase, and applying different annular dark-field rings. While the various STEM contrasts exhibit modulation following the periodicity of the standing light field, additional features appear along directions perpendicular to the light wavevectors. These variations arise from local morphological inhomogeneities of the FIB-prepared VO$_2$ lamella (Fig.~\ref{figlam}), such as thickness variations or bending, rather than from the optical illumination. Recent studies have suggested, that virtual dark-field imaging might be directly sensitive to strain fields \cite{gammer_diffraction_2015}. 

Figure~\ref{fig3}b highlights the contrast enhancement achieved with 4D STEM virtual dark-field imaging compared to conventional dark-field TEM, where a physical aperture selects a single diffraction spot. In virtual imaging, multiple diffraction spots can be selected simultaneously through post-processing rather than a single mechanical aperture, resulting in a higher signal-to-noise ratio (SNR) at a lower effective electron dose. For the comparison shown here, the STEM dwell time was adjusted such that the total electron dose over a given field of view was of the same order of magnitude as in the corresponding TEM dark-field measurements. The TEM images were additionally down-sampled to match the effective spatial resolution of the STEM data, and comparable fields of view were selected. Under these conditions, the comparison emphasizes the qualitative robustness of the virtual dark-field contrast rather than serving as a quantitative performance benchmark, which is beyond the scope of this work. 

The SNR can be further improved during post-analysis, for example by subtracting diffuse scattering contributions. Virtual apertures also eliminate the need for complex, sample-specific dark-field aperture arrays designed to collect multiple diffraction spots simultaneously, as employed in Ref.~\cite{danz_ultrafast_2021-1}. While the overall contrast trends remain similar, conventional bright- or dark-field imaging is sensitive to local sample bending and tilts, which can shift regions of the specimen out of the selected diffraction condition. In contrast, virtual dark-field imaging in 4D STEM sums over all equivalent diffraction spots, making it largely robust against such distortions. Enhanced contrast at reduced electron dose is particularly important for time-resolved measurements, where repeated pump--probe cycles can otherwise lead to cumulative beam damage. By extracting more information from each electron interacting with the specimen, ultrafast 4D STEM reduces the total number of electrons required for imaging, thereby minimizing both electron- and laser-induced damage.

\subsection*{Correlative ultrafast imaging}
\begin{figure}[htbp]%
\centering
\includegraphics[width=0.9\textwidth]{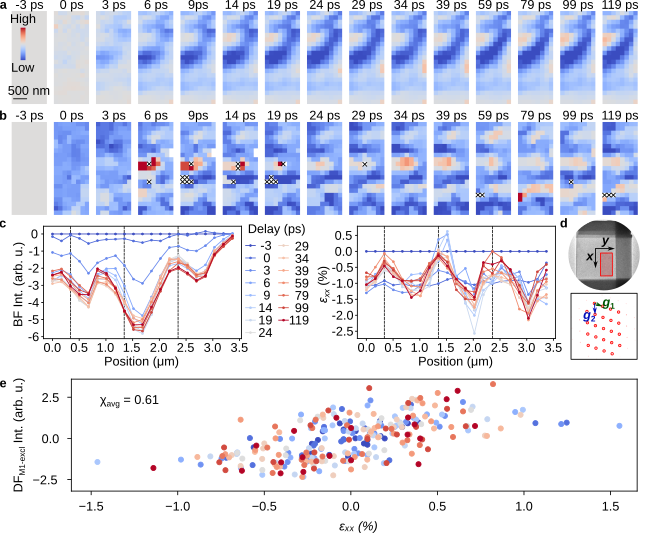}
\caption{\textbf{Bright-field imaging and strain analysis following transient grating excitation.} (a) Ultrafast bright-field imaging, showing contrast contributions from bending, diffraction, and other scattering effects. (b) Strain mapping from local diffraction pattern analysis, with black crosses indicating positions where strain retrieval was unsuccessful. (c) Integrated line profiles from virtual bright-field imaging, highlighting contributions from bending and diffraction contrast. The profiles are integrated along the y-direction. In both measurements, grating oscillations are clearly visible. (d) Measurement orientation in real and diffraction space. (e) Correlation plot of detrended virtual dark-field imaging (Fig. 3a panel 1) and strain indicating a positive correlation.}\label{fig4}
\end{figure}

Further, we compare bright-field imaging (Fig.~\ref{fig4}a) with strain mapping (Fig.~\ref{fig4}b) derived from 4D STEM. Strain imaging quantifies atomic displacements in reciprocal space, whereas bright-field imaging captures intensity variations in the direct (non-diffracted) beam arising from local changes in diffraction conditions, which can produce complex signals. In both cases, the laser-induced grating is clearly visible (Fig.~\ref{fig4}b). Bending effects, resulting from strain gradients, appear weaker near the sample edge where the lamella remains attached to the bulk, likely due to mechanical constraints suppressing out-of-plane deformation. In contrast, the in-plane strain components remain relatively uniform across the lamella. All images are normalized to negative time delays to account for static bending and strain conditions present prior to excitation. Other strain components are discussed in the SI.

To directly relate the observed lattice strain to the structural phase transition, we next quantify the correlation between the strain maps and Bragg-resolved virtual dark-field imaging sensitive to the monoclinic M1 phase (Fig.~\ref{fig4}c). Following laser excitation, we observe a correlation coefficient of approximately 0.6 between the M1-specific dark-field signal and the strain maps, indicating a strong positive correlation. The strain vector in our system is directly related to changes in the lattice parameter \textit{a}, which contracts by approximately 1\% during the phase transition \cite{kucharczyk_accurate_1979}. This confirms that the measured strain arises as a direct consequence of the structural phase transition. These findings underscore the value of quantitatively resolving strain evolution, providing clearer insights into the temporal and spatial coupling between structural transformations and lattice strain. Interestingly, strain and virtual bright-field imaging show no direct correlation; this is discussed in more detail in the SI.

To further support our conclusions, we performed finite element simulations (see SI for more information) in which a VO$_2$ lamella is heated by a transient laser grating, without including contributions from the phase transition. These simulations show that even at elevated temperatures, the resulting strain is insufficient to significantly shift the metal-to-insulator transition~\cite{park_measurement_2013} or to trigger the phase transition. The maximum simulated strain is roughly one order of magnitude smaller than the experimentally observed strain, indicating that thermal effects alone cannot account for the measured lattice deformation. In terms of timescale, the simulated strain field builds up over approximately 50~ps, whereas the experimental strain reaches its maximum after about 20~ps and remains constant thereafter. This faster rise reflects the combination of the ultrafast structural phase transition, which occurs on sub-picosecond timescales, and the subsequent mechanical redistribution of lattice deformation within the lamella. Together, the amplitude and temporal differences between simulation and experiment strongly support the conclusion that the measured strain is primarily a consequence of the structural phase transition, rather than laser-induced heating. Furthermore, the experimentally determined strain, on the order of 1\%, is too small to appreciably affect the phase transition, as it is expected to shift the transition temperature by less than $\sim$10~K~\cite{park_measurement_2013}, which is negligible under our experimental conditions.

\section*{Discussion}
In this work, we demonstrated the simultaneous mapping of strain and ultrafast imaging to capture a photo-induced propagating phase transition in a vanadium dioxide lamella. While strain does not initiate the transition in our case, it can play a critical role in modulating the transformation by locally lowering the energy barrier and influencing the transition temperature~\cite{aetukuri_control_2013}. By resolving quantitative strain dynamics with nanometer spatial and picosecond temporal resolution, we reveal how structural distortions emerge as a consequence of the phase transition. These insights enable rational strain-engineering strategies, such as controlling phase-nucleation sites, tuning transition thresholds, and stabilizing metastable phases through designed strain fields, which are critical for the development of ultrafast electronic components, phase-change materials, and next-generation energy-efficient switching devices.

We further demonstrated the creation of a transient transmission grating on picosecond timescales driven by the photoinduced phase transition and the accompanying refractive-index modulation. Through band-gap engineering - via strain, doping, or compositional tuning - this mechanism can in principle be shifted into the telecom wavelength range (1550 nm), where low-loss optical fiber communication operates. In this regime, the optically written transient gratings could function as ultrafast, reconfigurable diffraction elements, enabling spectral filtering, optical switching, and dynamic beam steering. More broadly, this approach establishes a materials-centric route to integrated photonic devices in which light propagation is actively controlled through phase-transition-driven structural and optical responses.

Additionally, we showed that virtual 4D STEM imaging achieves a better signal-to-noise ratio at a lower electron dose than conventional dark-field TEM for equivalent diffraction-based contrast. This reduced dose is particularly important for time-resolved measurements, where many pump–probe repetitions are required and cumulative beam damage can otherwise limit the achievable data quality. Looking ahead, we envision 4D STEM as a key technique in ultrafast transmission electron microscopy for probing structural and electronic dynamics during photo-induced phase transitions. While its first applications are still emerging, we anticipate significant advancements adapted from static 4D STEM. Improvements in direct-electron detectors and hardware synchronization will push the technique toward atomic resolution, as this regime requires not only single-electron sensitivity but also large dynamic range, and high detection efficiency. Further developments, such as electron ptychography and center-of-mass analysis, could enable full ultrafast phase imaging, including direct visualization of transient charge redistribution and other electronic degrees of freedom—an achievement that remains highly challenging with current off-axis electron holography on ultrafast timescales~\cite{houdellier_optimization_2019}. Future progress will depend on the development of more coherent pulsed electron sources with higher electron flux, improved direct-electron detectors, and enhanced mechanical stability. These advancements will continue to expand the capabilities of ultrafast 4D STEM, opening new possibilities for high-speed, high-resolution imaging in condensed matter physics and materials science.\\

\section*{Methods}

\subsubsection*{Sample Preparation}

Single crystals of VO$_2$ were grown by isothermal flux evaporation from 99.99\% V$_2$O$_5$ powder (Sigma-Aldrich) in a flowing nitrogen atmosphere at 1000~$^\circ$C, following established procedures~\cite{sasaki_new_1964,aramaki_single-crystal_1968}. Thin lamellas were prepared from VO$_2$ needles using focused ion beam (FIB) milling, yielding specimens with lateral dimensions of approximately $4 \times 5~\mu\text{m}^2$ and a thickness of $\sim150$~nm. The lamellas were mounted on a Gatan double-tilt holder for transmission electron microscopy experiments. Sample thickness was independently verified using electron energy-loss spectroscopy (EELS).

\subsubsection*{Ultrafast Electron Microscopy and Pump--Probe Configuration}

Ultrafast electron diffraction and 4D-STEM experiments were performed at the Ultrafast Electron Microscopy (UEM) laboratory at the KTH Royal Institute of Technology (Stockholm, Sweden). The experiments were conducted at room temperature in a pump--probe configuration. Photoelectron probe pulses with a full width at half-maximum (FWHM) duration of approximately 1.5~ps were generated and temporally characterized using photon-induced near-field electron microscopy (PINEM).

The optical pump pulses were derived from the same laser source (Tangerine, Amplitude Systemes), with a photon energy of $\sim1.2$~eV and a pulse duration of 300~fs (FWHM). The pump beam was focused to a spot size of approximately 200~$\mu$m on the sample. A 35$^\circ$ slanted aluminum-coated mirror with a reflectivity of around $50$\% was positioned above the sample plane to generate a transient optical grating via interference between the direct pump beam and its reflection. This configuration produced a sinusoidal excitation pattern with a spatial periodicity of 1~$\mu$m.

The relative time delay between the pump and probe pulses was controlled using a motorized delay stage (Newport ESP301). Experiments were performed at a repetition rate of 12~kHz, ensuring complete relaxation of the sample between successive excitation cycles.

\subsubsection*{Ultrafast 4D-STEM Acquisition}

In ultrafast 4D-STEM measurements, a focused electron probe was rastered across the sample, and a full diffraction pattern was recorded at each real-space position and pump--probe delay. This approach enables simultaneous acquisition of spatially resolved diffraction data and virtual imaging modes. Diffraction patterns were recorded using a hybrid pixel detector (CheeTah T3, Amsterdam Scientific Instruments), providing high sensitivity and dynamic range.

For 4D-STEM scans, a condenser aperture with a diameter of 50~$\mu$m was used to generate a quasi-parallel electron beam with a convergence semi-angle of a few milliradians. This configuration enabled clear separation of individual diffraction peaks, facilitating Bragg-resolved virtual imaging and quantitative strain analysis. The electron beam brightness was optimized to balance diffraction-space resolution and real-space imaging performance while minimizing electron dose.

Further details on the ultrafast electron pulse characterization, 4D-STEM acquisition parameters, strain analysis, and numerical simulations are provided in the Supporting Information.

\section*{Acknowledgment}
This research was funded by the Knut and Alice Wallenberg Foundation (2012.0321 and 2018.0104), the Swedish Research Council (VR), and through the ARTEMI national infrastructure (VR 2021-00171 and Strategic Research Council (SSF) RIF21-0026). A.N. acknowledges funding from the Swiss National Science Foundation (SNSF) through project P500PT\_214469. J. Wu gratefully acknowledges the Chinese Scholarship Foundation (CSC) for a doctoral fellowship.\\

\section*{Supporting information}
Sample preparation, COMSOL simulations of strain from laser-induced heating, Ultrafast transmission electron microscopy, Ultrafast 4D STEM data analysis, Ultrafast strain mapping, Time0 determination

\section*{Author Contributions}
A.N., I.K., and J.W. conceived the experiments. B.F. grew the vanadium dioxide crystals. A.N. conducted the experiments and performed the analysis. J. Wu performed the COMSOL simulations. I.K. and J.W. supervised the project. All authors contributed to the review and editing of the manuscript.

%\begin{comment}

\newpage
\section*{Supplementary information}

\subsection*{Correlative Ultrafast Imaging of a Propagating Photo-Driven Phase Transition Using 4D STEM}

Arthur Niedermayr,$^{1,*}$ Jianyu Wu,$^{1}$ Bertina Fisher,$^{2}$ Ido Kaminer,$^{3}$ and Jonas Weissenrieder$^{1,*}$

\medskip

$^{1}$Department of Materials and Nano Physics, School of Engineering Sciences,  
KTH Royal Institute of Technology, SE-100 44 Stockholm, Sweden

$^{2}$Department of Physics, Technion--Israel Institute of Technology,  
Haifa 32000, Israel

$^{3}$Department of Electrical and Computer Engineering,  
Technion--Israel Institute of Technology, Haifa 32000, Israel

\noindent
*~E-mail: arthurn@kth.se, jonas@kth.se

\subsection{Sample preparation }

Single crystals of VO$_2$, have been grown by isothermal flux evaporation \cite{sasaki_new_1964,aramaki_single-crystal_1968}  from 99.99\% V$_2$O$_5$ powder (Sigma Aldrich), in an atmosphere of flowing nitrogen at 1000°C. A single as-grown VO$_2$, needle, free of cracks and voids, was chosen for further processing with a dimension of about \( 0.05 \times 0.05 \times 3 \, \text{mm}^3 \).\\

\begin{figure}[htbp]%
\centering
\includegraphics[width=1\textwidth]{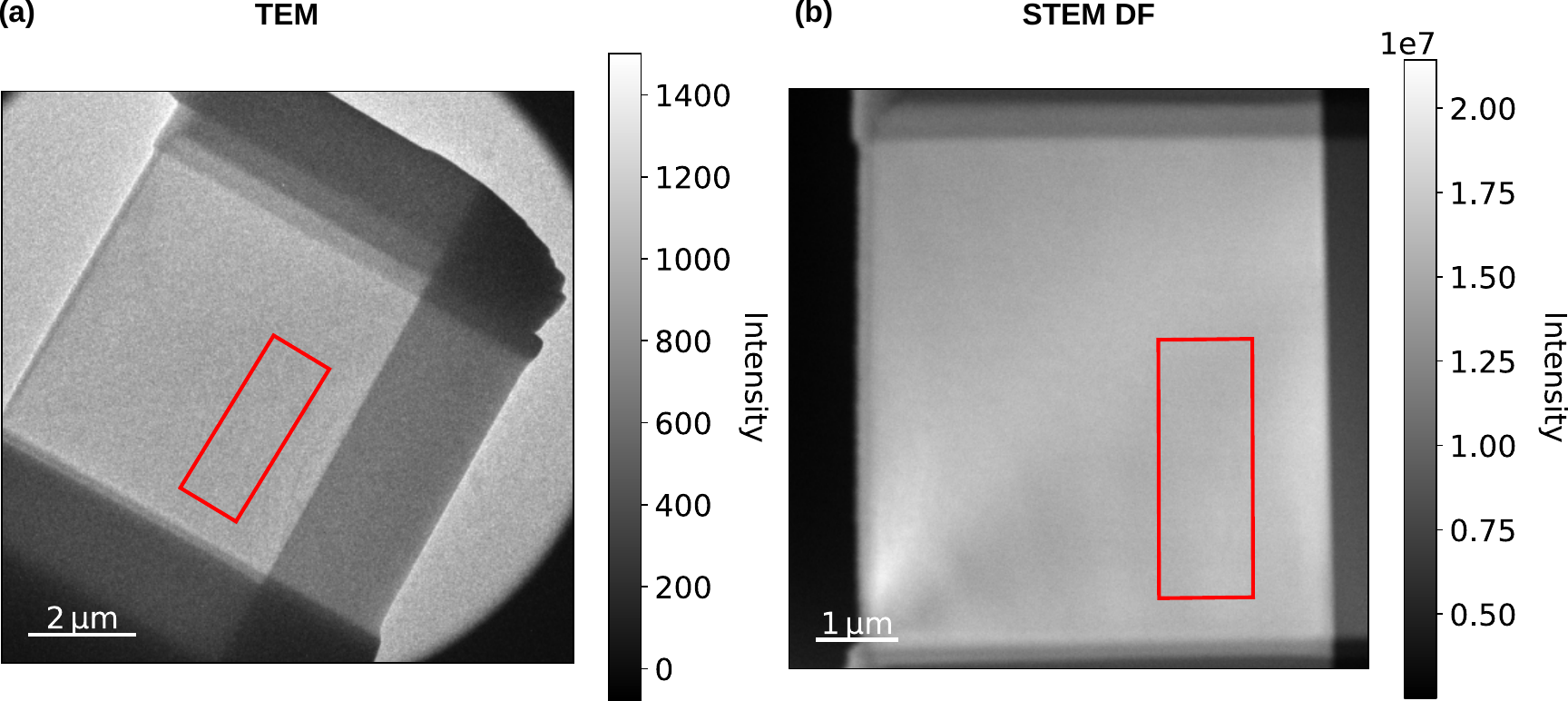}
\caption{\textbf{VO$_2$ lamella.} (a) TEM, (b) STEM DF. The red rectangle indicates the area where ultrafast STEM measurements where performed.}\label{figlam}
\end{figure}

Focused ion beam (FIB) milling was performed on this VO\(_2\) needle to create a FIB lamella with dimensions of \(4 \times 5 \, \mu \text{m}^2\) and a thickness of approximately 150 nm (Fig.~\ref{figlam}). FIB milling was carried out using a Ga ion source operated at 30 kV. Final thinning was performed at low ion energies to minimize Ga implantation and surface damage.\\

\subsection{COMSOL simulations of strain from laser-induced heating}
\label{comsol}

\begin{figure}[htbp]
    \centering
    \includegraphics[width=1\linewidth]{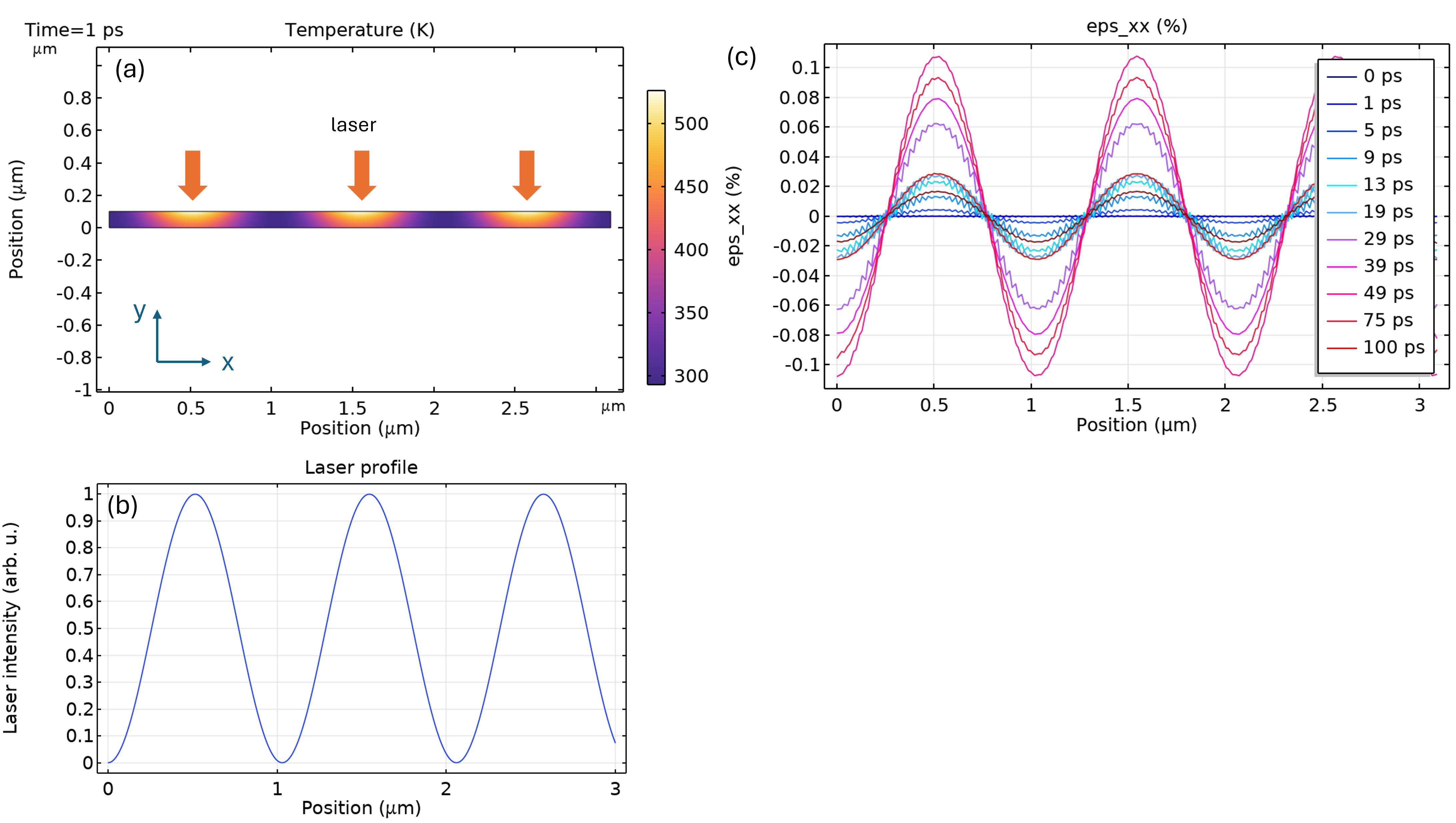}
\caption{\textbf{COMSOL simulations.} 
(a) Temperature distribution of a vanadium dioxide lamella heated by a transient laser grating. 
(b) Laser intensity profile used in (a) to heat the sample. 
(c) Simulated strain component $\varepsilon_{xx}$ resulting from laser-induced heating. No phase transition was included in this model, so it does not contribute to the strain. The simulated strain values remain an order of magnitude smaller than those observed experimentally.}
\label{figcom}
\end{figure}

To rule out strain as the primary driver of the structural phase transition, we performed finite element simulations (using the COMSOL softare package) of laser-induced heating in the monoclinic phase of a VO$_2$ lamella. Even under exaggerated excitation conditions—approximately twice the experimental laser fluence and with no phase transition explicitly modeled—the resulting strain reaches only 0.1\% after 50~ps, an order of magnitude smaller than the strain observed in the experiment. In these simulations, the VO$_2$ lamella reaches a maximum temperature of around 500~K (Fig.~\ref{figcom}a), which exceeds the bulk phase transition temperature of VO$_2$ at $\sim$340~K. The associated laser profile is shown in Fig.~\ref{figcom}b, and the resulting strain evolution confirms that the heating-induced strain remains negligible. If a lower pump intensity were used, the maximum temperature and the resulting strain would be reduced further. These results therefore provide a conservative, upper-limit estimate, indicating that the observed strain is not caused by laser-induced heating and is highly unlikely to trigger the phase transition. Moreover, literature reports suggest that significantly higher strain values would be required to meaningfully shift the transition temperature in vanadium dioxide~\cite{park_measurement_2013}.

The pump laser was modeled as a 300~fs pulse applied at $t = 0$ ps, with a volumetric power of $2.2 \times 10^{21}$~W/m$^3$. The laser power along the $x$-axis was modulated sinusoidally to simulate a transient optical grating (TOG). Light absorption in the thickness direction was modeled using the Beer-Lambert law. The sample was treated as free-standing, with periodic boundary conditions applied to the left and right sides. VO$_2$ material properties were taken from the COMSOL material library.

\textbf{The following parameters were used for the COMSOL simulations:}
\begin{itemize}
    \item Sample thickness: 100~nm
    \item Absorption coefficient: $5.9429 \times 10^4$~cm$^{-1}$~\cite{polyanskiy_refractiveindexinfo_2024}
    \item Heat capacity at constant pressure: 657~J/(kg$\cdot$K)~\cite{kizuka_temperature_2015}
    \item Thermal conductivity: 4.2~W/(m$\cdot$K)~\cite{kizuka_temperature_2015}
    \item Elasticity matrix (Voigt notation):~\cite{dong_elastic_2013}
\[
\mathbf{D} =
\begin{bmatrix}
D_{11} & D_{12} & D_{13} & D_{14} & D_{15} & D_{16} \\
D_{12} & D_{22} & D_{23} & D_{24} & D_{25} & D_{26} \\
D_{13} & D_{23} & D_{33} & D_{34} & D_{35} & D_{36} \\
D_{14} & D_{24} & D_{34} & D_{44} & D_{45} & D_{46} \\
D_{15} & D_{25} & D_{35} & D_{45} & D_{55} & D_{56} \\
D_{16} & D_{26} & D_{36} & D_{46} & D_{56} & D_{66} 
\end{bmatrix}
=
\begin{bmatrix}
418 & 210 & 156 & 0 & -10 & 0 \\
210 & 343 & 231 & 0 & 29 & 0 \\
156 & 231 & 394 & 0 & -33 & 0 \\
0 & 0 & 0 & 222 & 0 & 59 \\
-10 & 29 & -33 & 0 & 104 & 0 \\
0 & 0 & 0 & 59 & 0 & 169
\end{bmatrix} \times 10^9~\text{Pa}.
\]
\end{itemize}

\subsection{Ultrafast transmission electron microscopy}

The time-resolved three-dimensional experiments were performed in an ultrafast transmission electron microscope (a modified JEOL TEM 2100) operated at 200 kV, using a hybrid pixel detector (CheeTah T3, Amsterdam Scientific Instruments). The sample was excited with a 1030 nm wavelength laser (Tangerine, Amplitude Systems) with a 300 fs pulse duration. The pump laser was focused to a 200 $\mu$m (FWHM) spot with an average incident fluence of 5 mJ/cm$^2$. A 35-degree slanted, aluminum-coated mirror with a reflectivity of approximately 50\% was placed above the sample plane to generate a transient laser grating on the sample. The sinusoidal grating pattern with a periodicity of 1 $\mu$m arose from the interference between the direct pump beam and its reflection from the slanted face of the mirror. The vanadium dioxide lamella was excited at a repetition rate of 12 kHz, ensuring that the sample returned to its monoclinic ground state between subsequent laser pulses. The specimen was mounted on a double-tilt holder, allowing precise adjustment of the tilt angles $\alpha$ and $\beta$.\\

Electron probe pulses were generated through photoemission from a guard ring LaB\(_6\) cathode excited by 258 nm laser pulses. Further details of the experimental setup can be found in Ref. \cite{ji_influence_2017}. The temporal width of the electron bunches was approximately 1.5 ps, retrieved from photon-induced near-field electron microscopy measurements. The time delays between the pump-probe pulses were controlled using a motorized delay stage between the pump and probe beam path. \\

For the 4D STEM scans, we used the second smallest condenser aperture, with a diameter of 50 $\mu$m, to generate a more parallel electron beam with a convergence angle of a few mrad. This configuration allowed us to separate individual diffraction peaks. The electron beam brightness was fine-tuned to optimize the balance between resolution in diffraction space and imaging. The total exposure time per delay and pixel was 30 s, with an average electron flux of 100,000 electrons per 30 seconds incident on the direct electron detector. Experiments were performed at room temperature, with measurements taken along one of the main zone axes in the [01-1] projection of the monoclinic structure.\\

\begin{figure}[htbp]
\centering
\includegraphics[width=0.5\textwidth]{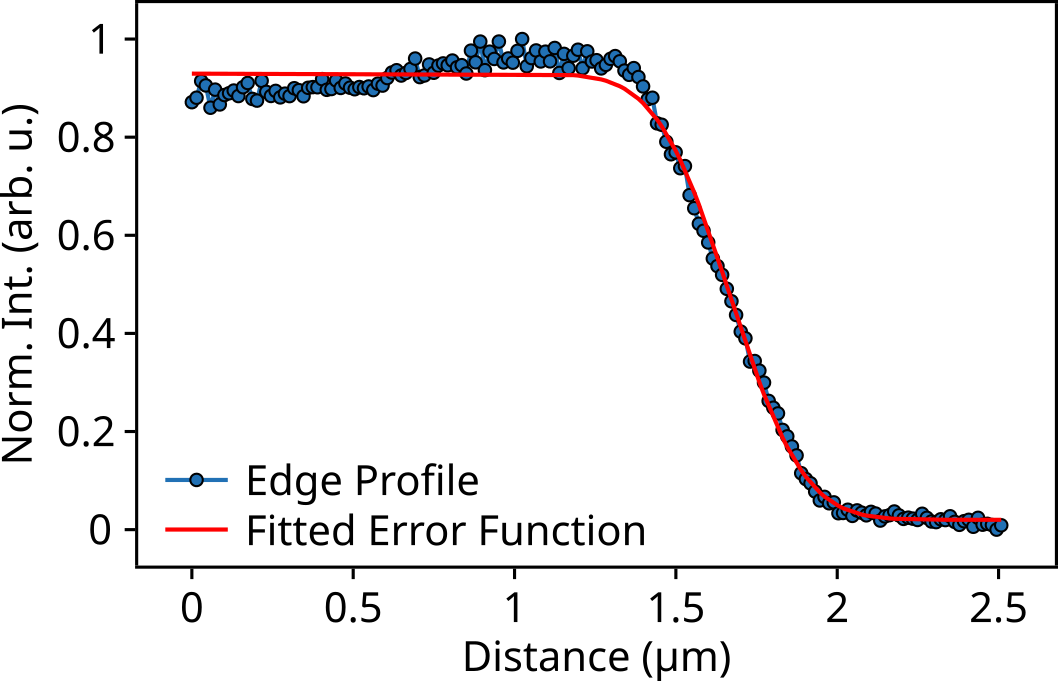}
\caption{\textbf{Beam size estimation from an error fit to the edge spread function.}}\label{fig5}
\end{figure}

The dataset consists of 21 × 8 = 168 measurement points in real space, with each point containing a 2D diffraction pattern acquired with a 30-second exposure time. A total of 15 4D STEM sets were collected, corresponding to 15 time delays, leading to a total acquisition time of approximately 21 hours. The nominal spatial resolution was 160 nm, determined by the pixel spacing, though the actual resolution was limited by the electron beam spot size. The effective beam spot size on the specimen was estimated by comparing a sharp specimen image with a reconstructed virtual STEM image. The electron beam spot size, determined from an error function fit to the edge spread function of a sharp edge, was approximately 400 nm (Fig. \ref{fig5}).\\

\subsection{Ultrafast 4D STEM data analysis}
The experimental data was analyzed as follows: For each time delay and each pixel in real space, a corresponding diffraction pattern was obtained. Background subtraction was performed using the difference of Gaussians method using the Pyxem software \cite{cautaerts_free_2022} to remove the inelastic background and thermal diffuse scattering near the zero-order diffraction peak. The diffraction patterns were also corrected for salt-and-pepper noise. To ensure consistency across time delays, the total intensity of all counts in real space was normalized, such that the sum of counts over all pixels was conserved across all time delays. All diffraction patterns were aligned with respect to each other.  Time-resolved virtual images were generated by applying a virtual mask to the diffraction patterns in Py4DSTEM \cite{savitzky_py4dstem_2021}  across all pixels in real space and for all time delays.\\

\begin{figure}[htbp]
\centering
\includegraphics[width=0.9\textwidth]{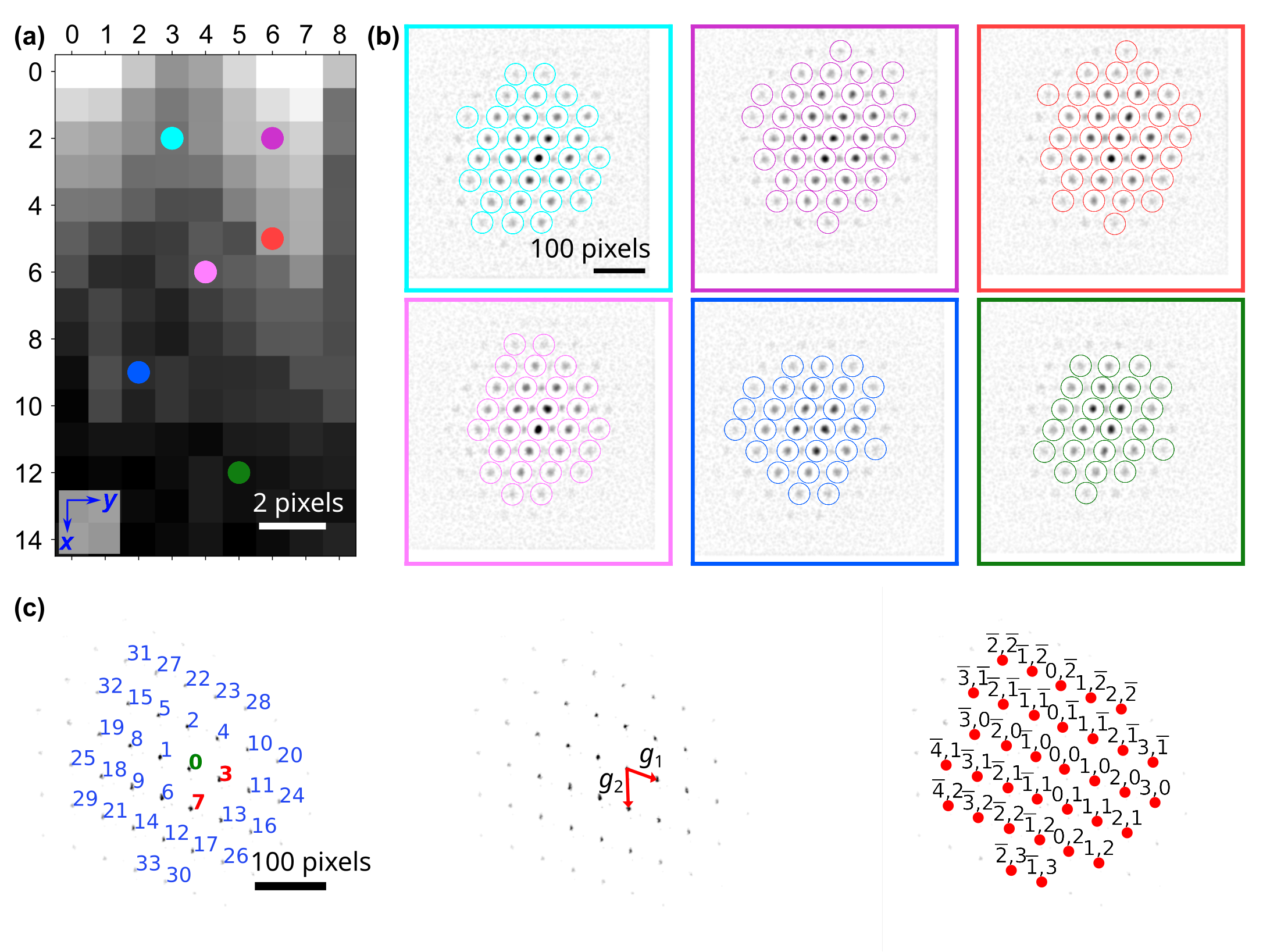}
\caption{\textbf{Strain analysis procedure using ultrafast nanobeam electron diffraction. }(a) Several pixels of the virtual image for a given time delay are selected to generate a template for strain analysis. The corresponding diffraction patterns along the [01-1] zone axis are shown in (b). Diffraction peaks are identified for strain analysis. The M1-exclusive diffraction spots are disregarded due to their low intensity. (c) The diffraction peaks are labeled, and the reciprocal lattice vectors are chosen. Deviations from these reciprocal lattice vectors indicate strain.}\label{fig6}
\end{figure}

\subsubsection{Raw data}
\begin{figure}[htbp]
    \centering
    \includegraphics[width=0.7\linewidth]{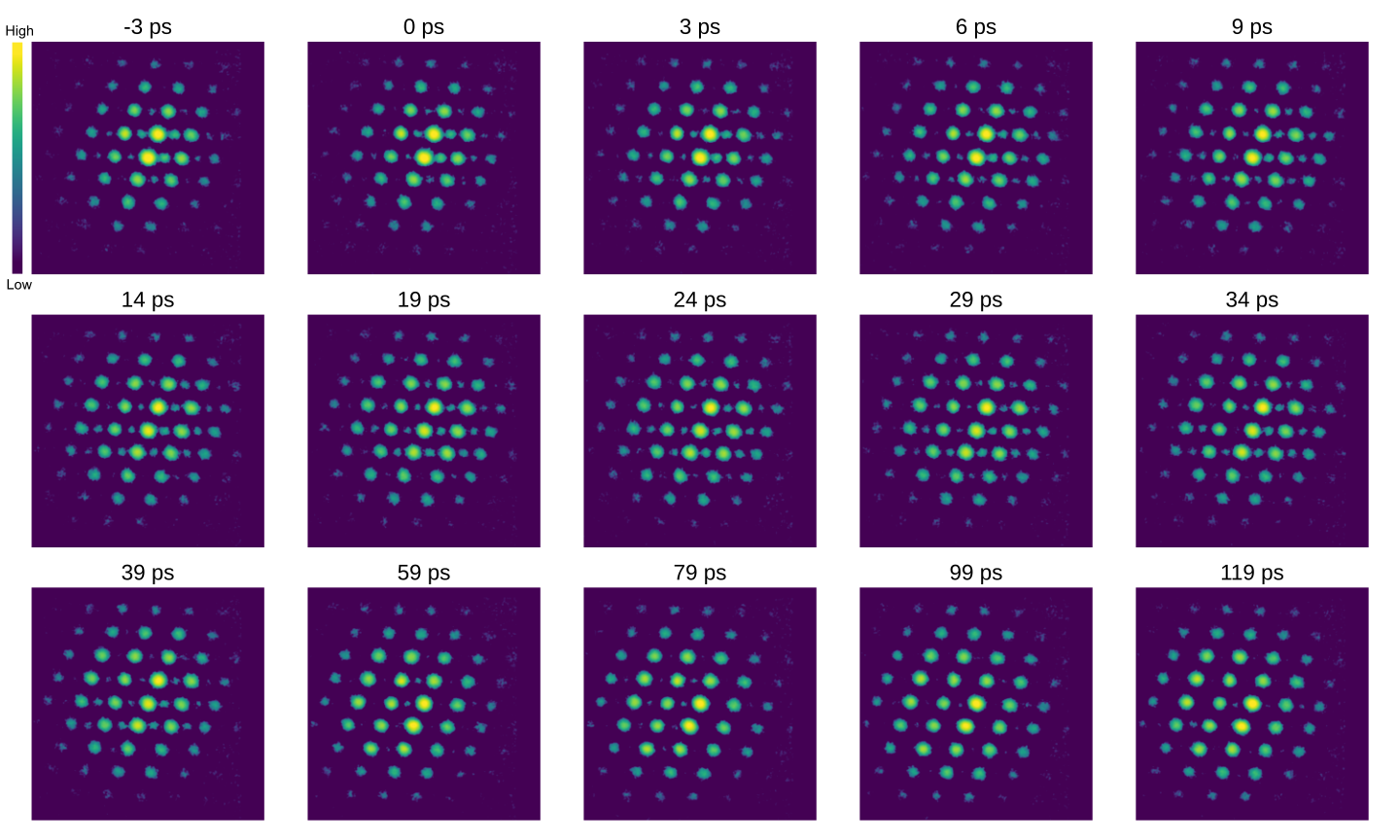}
\caption{\textbf{Ultrafast 4D STEM.} 
Diffraction patterns as a function of the pump-probe delay are shown for a single position in real space.}
\label{rawdata}
\end{figure}

Figure~\ref{rawdata} presents representative nano-beam electron diffraction patterns recorded at a fixed real-space position for increasing pump–probe delays. With increasing delay, the M1 superstructure reflections progressively weaken and eventually disappear, consistent with the photo-induced suppression of the monoclinic lattice distortion.

\subsection{Ultrafast strain mapping}
Strain analysis was conducted using the py4DSTEM software \cite{savitzky_py4dstem_2021} . The strain at different time delays was referenced to negative delays by using the same reciprocal lattice vectors. At negative delays, the total strain was set to zero following a median strain approach. The work flow of strain analysis is shown in Fig. \ref{fig6}. Due to symmetry considerations of the structured light excitation, we analyze in the main text only the strain along the x-direction, $\varepsilon_{xx}$, which is perpendicular to the generated optical grating fringes.

\subsubsection{Other strain components}
\label{sec:strain_yy}
\begin{figure}[htbp]
\centering
\includegraphics[width=0.8\textwidth]{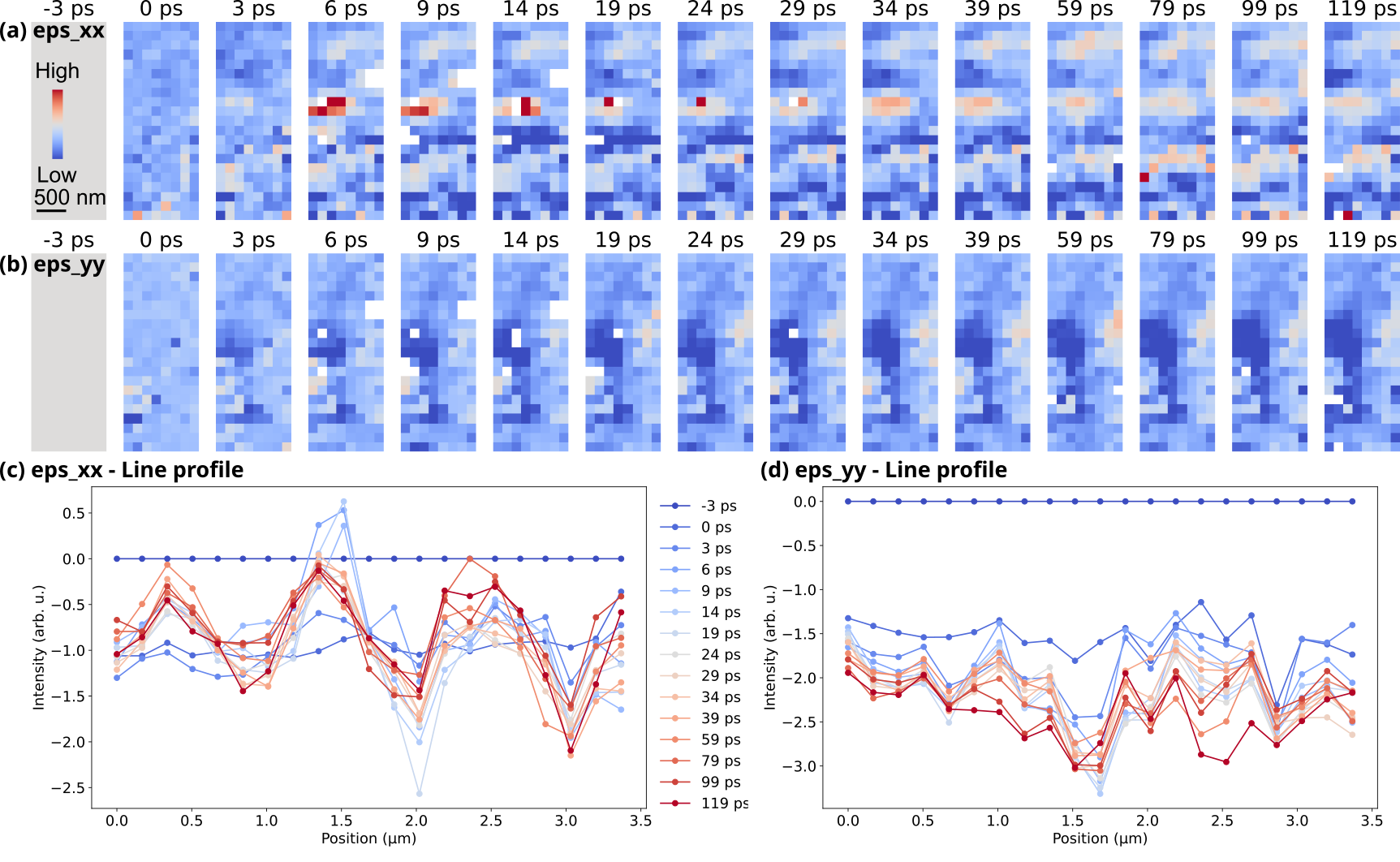}
\caption{\textbf{Strain tensor components $\varepsilon_{xx}, \varepsilon_{yy}$ and the corresponding line profiles. }}\label{figSIstrain}
\end{figure}

To give a complete analysis of the strain occurring during the phase transition, we present here additionally the strain component $\varepsilon_{yy}$ (Fig.~\ref{figSIstrain}). However, since there is no transient grating modulation along this direction, this strain component doesn't exhibit clear features.

\subsubsection{Correlation of strain with virtual BF and DF imaging}
\label{sec:corrBF}
\begin{figure}[htbp]
\centering
\includegraphics[width=0.7\textwidth]{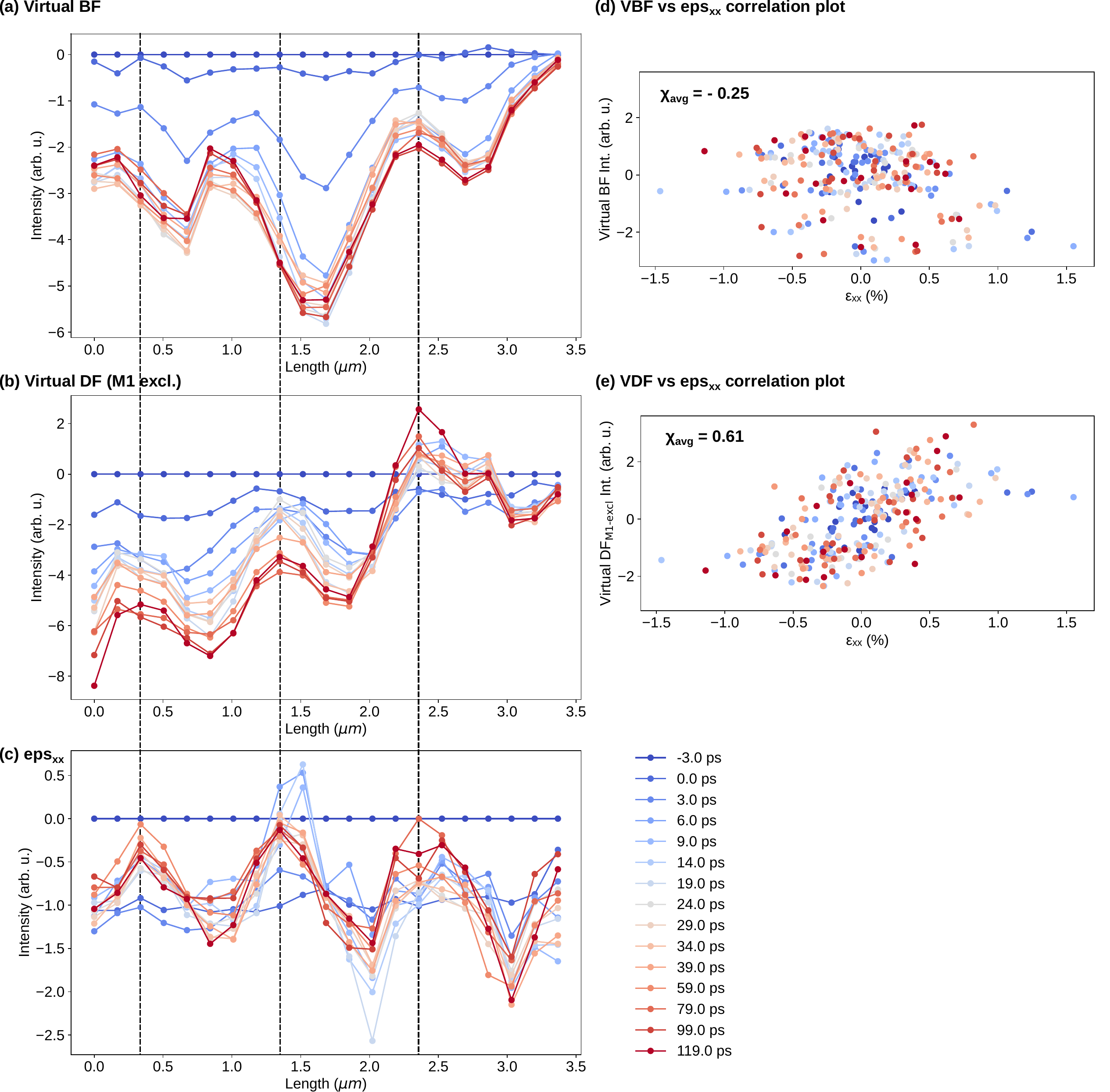}
\caption{\textbf{Comparison of line profiles of virtual bright-field, virtual dark-field and $\mathbf{\varepsilon_{xx}}$ imaging and their correlations.} All quantities were extracted from a single ultrafast 4D STEM scan, ensuring identical temporal sampling and spatial registration across the virtual imaging modes and the strain analysis.}\label{fig_correlation}
\end{figure}

Fig.~\ref{fig_correlation} compares line profiles of virtual bright-field (VBF), virtual dark-field (VDF), and the extracted strain $\varepsilon_{xx}$, along with their correlations.

In Fig.~\ref{fig_correlation}e, the strong positive correlation between the VDF signal associated with the M1 Bragg reflections and the extracted strain ($\chi \approx 0.61$) indicates that both observables probe the same underlying structural order parameter. The VDF intensity directly reflects the presence and strength of the monoclinic lattice distortion, while the strain maps quantify the accompanying lattice deformation. Their in-phase spatial modulation therefore confirms that the measured strain is intimately linked to the structural phase transition.

In contrast, Fig.~\ref{fig_correlation}d shows the weak and negative correlation between the VBF signal and the strain ($\chi \approx -0.25$), highlighting the fundamentally different contrast mechanism of bright-field imaging. VBF contrast is sensitive to a convolution of effects, including diffuse scattering, local lattice disorder, and sample bending or tilts. These contributions can respond differently to the structural distortion associated with the phase transition, leading to a reduced and phase-shifted correlation with the strain. As a result, although VBF also exhibits spatial modulation following the transient grating excitation, its contrast is not phase-locked to the structural order parameter.

Taken together, these correlation values quantitatively demonstrate that diffraction-based virtual dark-field imaging and strain mapping provide selective, structure-sensitive information, whereas bright-field contrast mixes multiple scattering channels and is therefore less suitable for directly tracking the microscopic coupling between strain and the structural phase transition.

\subsection{Time0 determination}
\begin{figure}[htbp]
\centering
\includegraphics[width=0.6\textwidth]{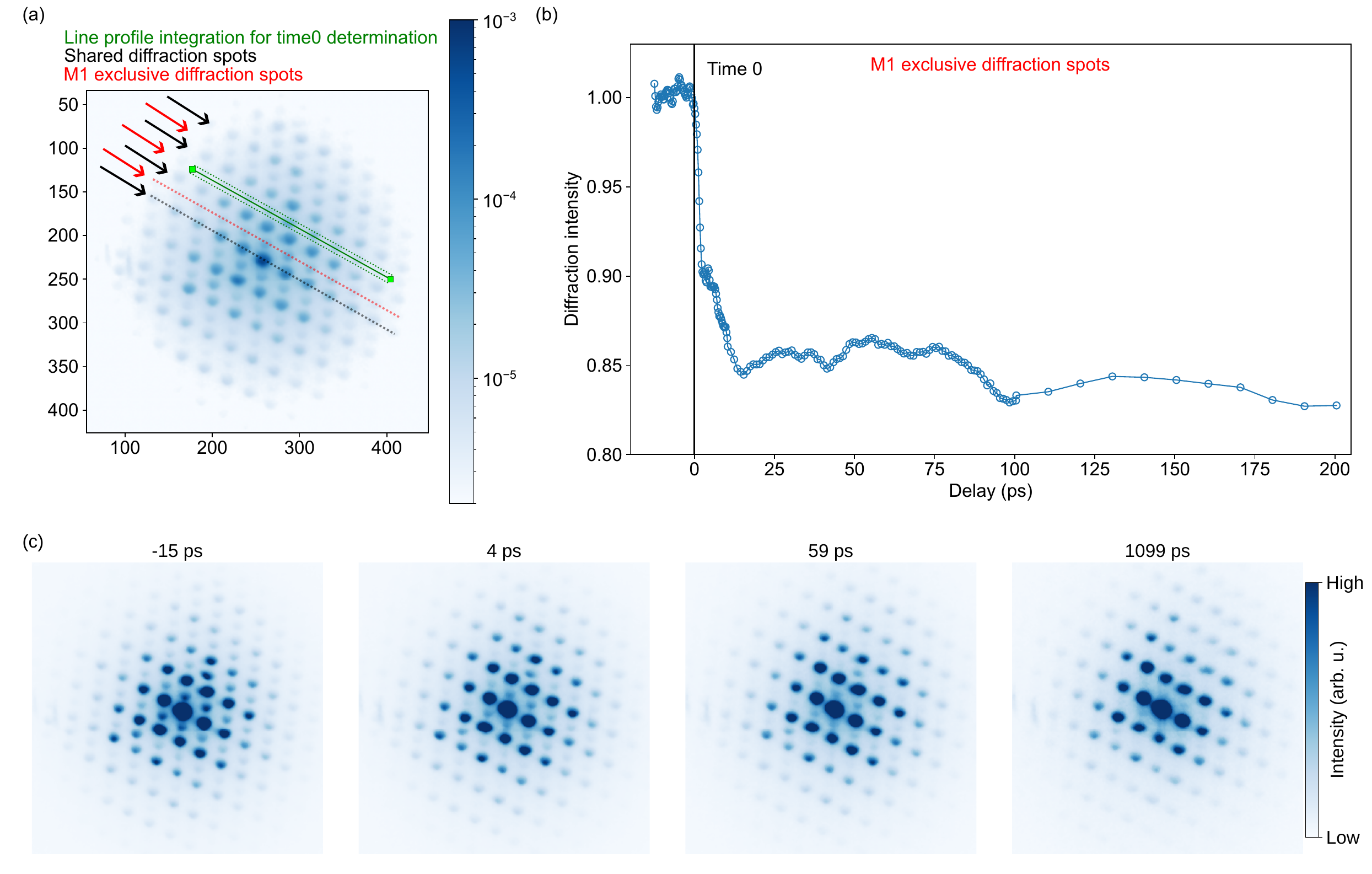}
\caption{\textbf{Experimental determination of time0.} (a) Diffraction pattern of  VO$_2$ in the monoclinic phase. (b) Integration over a line profile of M1-exclusive diffraction spots allows for a robust determination of time0 in an ultrafast transmission electron microscope. (c) Electron diffraction patterns at indicated time delays.}\label{figtime0}
\end{figure}

In our experiments, the time0 between the laser pump and electron probe is determined by monitoring the intensity of diffraction spots that are exclusive to the M1 monoclinic phase of VO$_2$. By integrating the intensity of these M1-specific diffraction peaks as a function of pump-probe delay (Fig.~\ref{figtime0}), the time at which the signal begins to change indicates the temporal overlap of the pump and probe pulses. This approach provides a reference for the onset of photoinduced structural dynamics, as the reduction in intensity of M1-exclusive spots corresponds directly to the transition toward the rutile phase.

\newpage
%\end{comment}

\bibliography{lib_202411_01}% common bib file



\end{document}